\documentclass[a4paper]{article}
\usepackage{INTERSPEECH_v2}
\usepackage{float}
\title{Voice Conversion Using Sequence-to-Sequence Learning of\\
 Context Posterior Probabilities}
\name{Hiroyuki Miyoshi, Yuki Saito, Shinnosuke Takamichi, and Hiroshi Saruwatari}
\address{Graduate School of Information Science and Technology, The University of Tokyo,\\ 7-3-1 Hongo, Bunkyo-ku, Tokyo 113-8656, Japan}
\email{mathma1306@gmail.com\\ \{yuuki\_saito, shinnosuke\_takamichi, hiroshi\_saruwatari\}@ipc.i.u-tokyo.ac.jp
}

\renewcommand{\Vec}[1]{\textrm{\boldmath$#1$}} 
\newcommand{\x}{\Vec{x}} 
\newcommand{\y}{\Vec{y}} 
\begin{document}
\maketitle

\begin{abstract}
Voice conversion (VC) using sequence-to-sequence learning of
context posterior probabilities is proposed.
Conventional VC using shared context posterior probabilities
predicts target speech parameters from the context posterior probabilities estimated from
the source speech parameters. Although conventional VC can be
built from non-parallel data, it is difficult to convert speaker
individuality such as phonetic property and speaking rate contained in the posterior probabilities
because the source posterior probabilities
 are directly used for predicting target speech parameters.
 In this work, we assume that the training data partly include parallel speech data and
 propose sequence-to-sequence learning between the source and target posterior probabilities.
  The conversion models perform non-linear and variable-length
  transformation from the source probability sequence to the target one.
  Further, we propose a joint training algorithm for the modules.
  In contrast to conventional VC, which separately trains the speech recognition that estimates
  posterior probabilities and the speech synthesis that predicts target speech parameters,
  our proposed method jointly trains these modules along with the proposed probability conversion modules.
Experimental results demonstrate that our approach outperforms the conventional VC.
\end{abstract}

\vspace{-5pt}
\section{Introduction}
Voice conversion (VC) is a technique for converting para- and non-linguistic
information while keeping linguistic information.
VC is used in a variety of applications such as speech enhancement~\cite{Kain_dys,Rudzicz} and
language education for non-native speakers~\cite{aryal14}.
It is mainly classified into two types:~text-independent VC
and text-dependent VC.
Text-independent VC directly predicts target speech parameters
from the source speech parameters,
and acoustic models such as Gaussian mixture models~\cite{toda07_MLVC,YSty98}
or deep neural network~\cite{desai09}
are trained using only speech data.
Since the models are often trained using parallel speech data,
the conversion quality of the performance is typically highly accurate.
However,
in most cases,
parallel data is not readily available.
Text-dependent VC~\cite{kain98,sundermann06usvc},
in contrast,
converts speech parameters through textual information.
This type consists of two modules:
speech recognition that estimates the textual information from the source speech,
and
speech synthesis that predicts target speech from the textual information.
Basically,
parallel data are not required to build the VC,
and the training data are easily available.
However,
the conversion units of this method are rougher
(e.g., phoneme, word, or other linguistic units)
than those of text-independent VC (e.g., frame).
VC using shared context posterior probabilities~\cite{LSun16}
is classified in text-dependent VC,
but the conversion unit is frame level.
The context posterior probability of source speech parameters is
estimated frame by frame,
and then the target speech parameters are
predicted from the estimated posterior probabilities.
This VC is interpreted as {\it soft} text-dependent VC
and can be extended to cross-lingual text-to-speech~\cite{LSun16_2,Xie+2016}.
However,
it cannot convert speaker individuality
(e.g., speaking rates and phonetic properties)
included in the context posterior probabilities
because the posterior probabilities of the source speech
are directly used for predicting target speech parameters.

In light of the above,
we propose a sequence-to-sequence learning of the
context posterior probabilities.
Assuming that the training data partly include
parallel speech data (parallel utterances of phrases),
we build an encoder-decoder model~\cite{EncDec}
that converts the posterior probabilities of the source speech parameters
into those of the target speech parameters.
The proposed posterior probability conversion module is inserted
between conventional speech recognition and synthesis.
When we do not build the conversion model or
do not have parallel data,
conventional VC~\cite{LSun16} is available.
Further,
we propose a joint training algorithm.
Whereas the conventional VC~\cite{LSun16} separately trains speech recognition
and speech synthesis,
our approach jointly trains these modules
(like auto-encoding) and the proposed conversion module.
We found through experiment that
the proposed methods outperform the conventional VC~\cite{LSun16}.

\vspace{-5pt}
\section{VC Using Shared Context Posterior Probabilities}
\begin{figure}[t]
  \begin{center}
  \includegraphics[width=0.875\linewidth,clip]{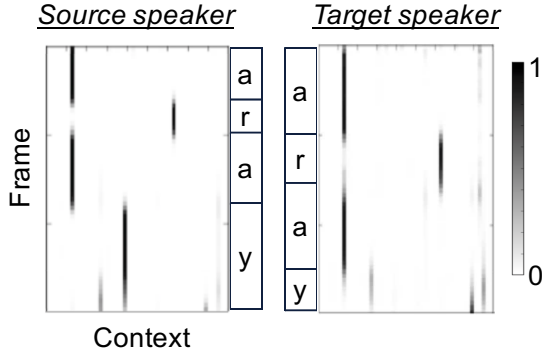}
  \caption{An example of source (left) and target (right) context posterior probabilities.}
  \label{ppg}
  \vspace{-20pt}
\end{center}
\end{figure}

Conventional VC using shared context posterior probabilities~\cite{LSun16}
contains two modules: speech recognition and speech synthesis.
They are separately trained,
and the voice conversion is performed by concatenating them.
Figure~\ref{ppg} shows an example of context posterior probabilities.
The upper and middle parts of Fig.~\ref{proposedmethod}
show the details of these processes.

\subsection{Training Stage}
Recognition models estimate context posterior probability sequence
from the speech parameter sequence.
Let
$\x = [\x_{1}^\top,\cdots,\x_{T_{x}}^\top]^\top$
and
$\y = [\y_{1}^\top,\cdots,\y_{T_{y}}^\top]^\top$
be source and target speech parameter sequences,
respectively.
$\x_t$ and $\y_t$
are the parameters at frame $t$.
$T_x$ and $T_y$ are their frame lengths.
Also,
let
$\Vec{l}^{(x)} = [\Vec{l}_{1}^{(x)},\cdots,\Vec{l}_{T_{x}}^{(x)}]^\top$
and
$\Vec{l}^{(y)} = [\Vec{l}_{1}^{(y)},\cdots,\Vec{l}_{T_{y}}^{(y)}]^\top$
be the context label sequence
(such as quin-phone)
corresponding to $\x$ and $\y$,
respectively.
Speaker-independent neural network
$\Vec{R}(\cdot)$
is trained using speech data including
$\x$ and $\y$,
and the training criterion is minimizing the cross entropy
$L_{C}(\Vec{l}^{(x)},\Vec{R}(\x))$.

Synthesis models predict target speech parameter sequence $\y$
from the corresponding context posterior probability sequence $\hat{\Vec{p}}_y$,
using the trained recognition models,
i.e.,
$\hat{\Vec{p}}_y = \Vec{R}(\y)$.
The target-speaker-dependent neural networks
$\Vec{G}(\cdot)$ are trained to minimize the mean squared error
$L_G(\y, \Vec{G}(\hat{\Vec{p}}_y))$
between
$\y$
and
$\Vec{G}(\hat{\Vec{p}}_y)$.

\subsection{Conversion Stage}
In conversion,
the converted speech parameter sequence $\hat{\y}$ is
predicted by concatenating speech recognition and speech synthesis,
i.e.,
$\hat \y = \Vec{G}(\hat{\Vec{p}}_x) = \Vec{G}(\Vec{R}(\x))$,
where
$\hat{\Vec{p}}_x$
is the context posterior probability sequence of $\x$.
Note that the frame lengths of
$\x$, $\hat{\Vec{p}}_x$ and $\hat{\Vec{y}}$ are the same,
i.e., $T_x$.

\subsection{Problems}
Since the posterior probabilities estimated in speech recognition
are directly used for speech synthesis,
it is difficult to convert speaker individuality included in the posterior probabilities,
such as the speaking rate (frame length) and phonetic characteristics
(see Fig.~\ref{ppg}).
Also,
improving recognition accuracy does not always improve speech quality
in converted speech
(except zero error in recognition).

\vspace{-5pt}
\section{Proposed VC using Sequence-to-Sequence Learning
of Context Posterior Probabilities}

To overcome the limitation of the conventional method,
we propose an approach for converting source context posterior probabilities
to target context posterior probabilities
using sequence-to-sequence learning.

\subsection{Sequence-to-Sequence Learning}
\begin{figure}[t]
  \includegraphics[width=0.875\linewidth,clip]{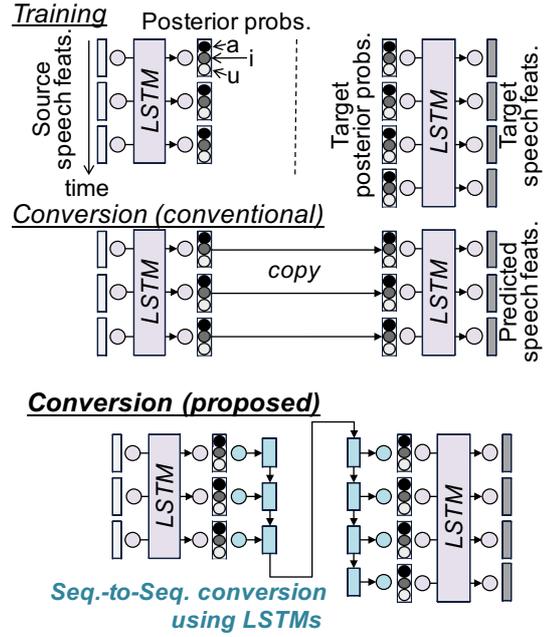}
  \caption{
  Training and conversion procedures of conventional and proposed VC.
  In the proposed VC,
  the source context posterior probabilities are transformed
  into the target posterior probabilities.
  }
  \label{proposedmethod}
  \vspace{-10pt}
\end{figure}

Sequence-to-sequence learning using recurent neural networks (RNNs)~\cite{Seq2Seq}
can be applied to the problem
that the source and target sequences have different lengths.
An encoder-decoder model we adopt here maps a variable-length
source sequence to a fixed-length vector,
and maps the vector to the variable-length target sequence.
At each frame,
the source side RNN (encoder) and target side RNN (decoder)
predict the source and target features of the next frame,
respectively.
As discussed below,
we adopt this in order to convert a source posterior probability sequence
to a target that has a different length.
The proposed procedure is shown in the lower part of Fig.~\ref{proposedmethod}.

\subsection{Training Stage}
We propose two algorithms to perform the posterior probability conversion.
The first algorithm {\it separately} trains speech recognition
and synthesis the same as the conventional algorithms
and {\it separately} trains probability conversion models
using the source and target posterior probabilities.
The second algorithm {\it jointly} trains the recognition and synthesis
and trains the conversion models considering
not only posterior probabilities conversion but also speech synthesis.
In conversion,
we concatenate
the three models for converting input speech parameters.

\subsubsection{Training of probability conversion models}
Given the parallel sequences of source and target context posterior probabilities,
we train encoder-decoder models
$\Vec{C}(\cdot)$
that convert the source and target sequences.
The loss function to be minimized is as follows:
\begin{align}
  L(\Vec{l}^{(y)},\hat{\Vec{p}}_{x},\hat{\Vec{p}}_{y}) = L_{G}(\hat{\Vec{p}}_{y},\Vec{C}(\hat{\Vec{p}}_{x})) + L_{C}(\Vec{l}^{(y)},\Vec{C}(\hat{\Vec{p}}_{x})).
  \label{eq_loss1}
\end{align}
The first term minimizes the conversion error
between the predicted and target sequences.
The second term minimizes the cross-entropy using $\Vec{l}^{(y)}$,
which was obtained by the training of $\Vec{R}(\cdot)$,
and can decrease the recognition error included in $\hat{\Vec{p}}_{y}$.
Our preliminary evaluation demonstrated that using this formulation
results in better conversion accuracy than using only the first term.

Sequence-to-sequence learning suffers from long-term dependencies,
i.e., error accumulation,
so in our approach,
we implement phoneme-by-phoneme probability conversion.
Given the phoneme boundary of the source and target probability sequence,
phoneme-independent encoder-decoder models are trained
to convert the probability sequence within the current phoneme.

\subsubsection{Jointly training of recognition, synthesis, and conversion}
Since the final goal of the method is to minimize synthesis error,
its pre-processes
(i.e., recognition and conversion)
must be trained by considering this error.
We train speech recognition $\Vec{R}(\cdot)$ to minimize
not only recognition error but also synthesis error
(e.g., reconstruction error of auto-encoders):
the loss function is
$L_{C}(\Vec{l}^{(x)},\Vec{R}(\x))+L_{G}(\x, \Vec{G}(\Vec{R}(\x)))$.
The speech synthesis $\Vec{G}(\cdot)$ is trained in the conventional manner.
We further train the conversion models to minimize not only conversion error
but also synthesis error:
the loss function is the sum of Eq.~(\ref{eq_loss1})
and
$L_{G}(\y, \Vec{G}(\Vec{C}(\hat{\Vec{p}}_x)))$.

\subsection{Discussion}
Text-dependent VC forcibly aligns the input speech feature segment
into a single context
(e.g., phoneme, syllable, or word unit)
and
generates output speech features from the context sequence.
Although this method can flexibly transform the context sequence
(e.g., variable-length conversion),
it cannot avoid the effect of time quantization
with mapping from speech feature segments.
Meanwhile,
text-independent VC with dynamic time warping (DTW)~\cite{toda07_MLVC}
aligns speech features in a frame level,
but it limits the transformation of
(implicitly considered) context sequence,
e.g., the sequence length is fixed.
The conventional method~\cite{LSun16,LSun16_2}
also corresponds to the latter
because
the source posterior probability is directly used
for synthesizing the target speech.
In comparison with these methods,
since the proposed algorithm performs
frame-level conversion without forced alignments,
it can avoid the effect of time quantization
and convert context sequence flexibly.

Joint training of recognition and synthesis,
which is proposed in Section 3.2.2,
is similar
to auto-encoding processes~\cite{Yoshua06}
with the referred class labels and dual learning~\cite{dual}.
Therefore,
we expect that these processes can be extended to
the supervised learning of variational auto-encoders~\cite{DPKMW14}
that have not only class labels
(e.g., context labels)
but also the hidden variables~\cite{Chin16,Die14}.

\vspace{-5pt}
\section{Experiments}
\subsection{Experimental Setup}
Although the conventional VC~\cite{LSun16} and proposed VC accept non-parallel
speech data and partly included parallel data,
we used fully parallel data in this evaluation.
We prepared speech data of eight speakers
taken from the ATR Japanese speech database~\cite{sagisaka90}.
The speaker uttered 503 phonetically balanced sentences.
We built the speaker-independent speech recognition module with speaker codes
by using the speech data of eight speakers including source and target speakers.
We built conversion and synthesis modules by using
speech data of source female and target male speakers.
We used 450 sentences (subsets A to I) for the training
and
53 sentences (subset J) for the evaluation.
Speech signals were sampled at a rate of 16~kHz,
and the shift length was set to 5~ms.
The 0th--through--24th mel-cepstral coefﬁcients were used
as a spectral parameter
and
$F_0$ and 5 band-aperiodicity~\cite{kawahara01,ohtani06}
were used as excitation parameters.
We used the STRAIGHT analysis-synthesis system~\cite{kawahara99}
for the parameter extraction and waveform synthesis.
To improve training accuracy,
speech parameter trajectory smoothing~\cite{takamichi15blizzard}
with a 50~Hz cutoff modulation frequency
was applied to the spectral parameters in the training data.
In the training phase,
spectral features were normalized to have zero-mean unit-variance,
and 80~\% of the silent frames were removed from the training data.
We used AdaGrad~\cite{adagrad} as the optimization algorithm,
setting the learning rate to 0.01.
Both in speech recognition and generation stage,
we used bi-directional long short-term memory (LSTM)~\cite{LSun16,LSun16_2}
and
each hidden layer contained 256 units.
For probability conversion,
we used bi-directional LSTM in encoder, and LSTM in decoder.
Each hidden layers contained 256 units.

Quin-phone was used as the context labels.
For training the recognition models,
we divided the quin-phone into five groups:
previous phoneme, current phoneme, and next phoneme, and so on.
The cross-entropy loss was calculated for each group,
and the loss function for training the recognition models was
the sum of each loss~\cite{huang15}.
Only spectral and their delta features were used for recognition and synthesis.
In the proposed method,
$F_0$ was linearly transformed~\cite{toda07_MLVC} first,
and we modified its length using DTW
between
the source context posterior probability sequence
and
the converted posterior probability sequence.
Only DTW was used for band-aperiodicity conversion.
This evaluation uses reference phoneme duration
for converting posterior probabilities
in order to address the sequence length determination problem
to which sequence-to-sequence learning is prone~\cite{errorofac}.
Given the phoneme duration of the target natural speech parameter sequence
in conversion data,
we performed phoneme-level probability conversion.
Given the phoneme duration of the source and target speech parameters
in the training and conversion data,
the conversion models converted the probabilities within the current phoneme.
The finally generated posterior probability sequence
was then calculated by concatenating the converted probabilities.

Two evaluations were performed to compare the conventional and proposed VC.
First,
we evaluated the effectiveness of the proposed posterior probability conversion,
and then,
we evaluated the effect of the proposed joint training algorithms.

\subsection{Evaluations}
We discuss the effectiveness
of the proposed posterior probability conversion.
The separately trained modules were used here.
\subsubsection{Objective Evaluation}
We calculated mel-cepstral distortion
between the target and converted speech parameters
of the conventional VC~\cite{LSun16} and proposed VC.
DTW was used to align the target and converted parameters
by the conventional VC.
The difference between the two methods
is the time warping method, i.e., DTW or sequence-to-sequence learning.
\begin{figure}[t]
 \begin{center}
 \includegraphics[width=0.9\linewidth]{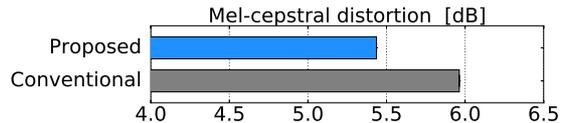}
 \caption{
  Mel-cepstral distortion of conventional VC
  and
  proposed VC integrating posterior probability conversion.
 }
 \vspace{-20pt}
 \label{MCD}
 \end{center}
\end{figure}
The results (Fig.~\ref{MCD}) clearly show that
the proposed VC outperforms the conventional VC,
we demonstrate that spectral distortion caused by DTW
can be alleviated by the use of sequence-to-sequence learning.

\subsubsection{Subjective Evaluation}
To subjectively evaluate the conventional and proposed VC,
we performed a preference AB test to evaluate the converted speech quality.
We presented every pair of converted speech of the two sets in random order
and
had listeners select the speech sample that sounded better.
Similarly,
we performed an XAB test on the speaker individuality
using the natural speech as a reference ``X.''
Seven listeners participated in each assessment.

\begin{figure}[t]
  \begin{center}
  \includegraphics[width=0.9\linewidth,clip]{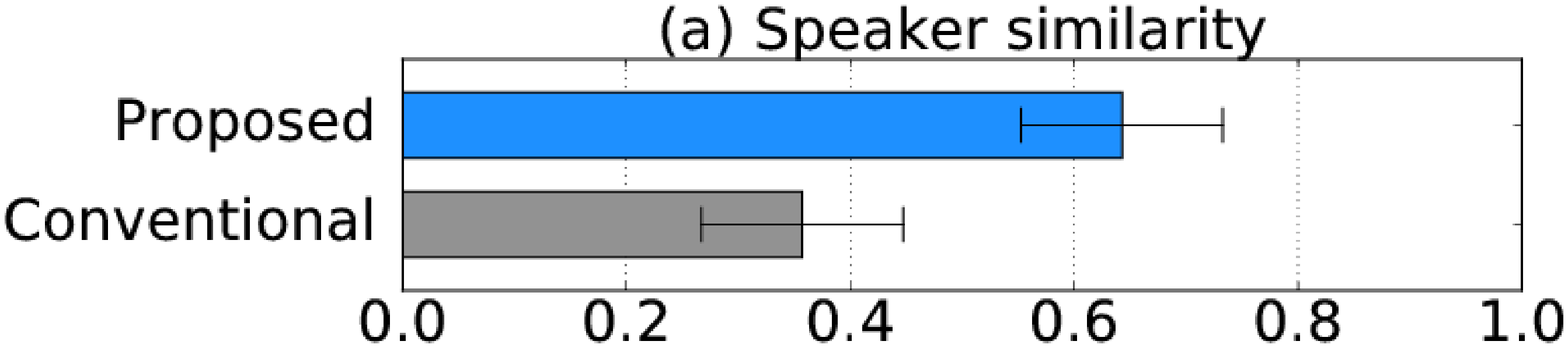}
  \includegraphics[width=0.9\linewidth,clip]{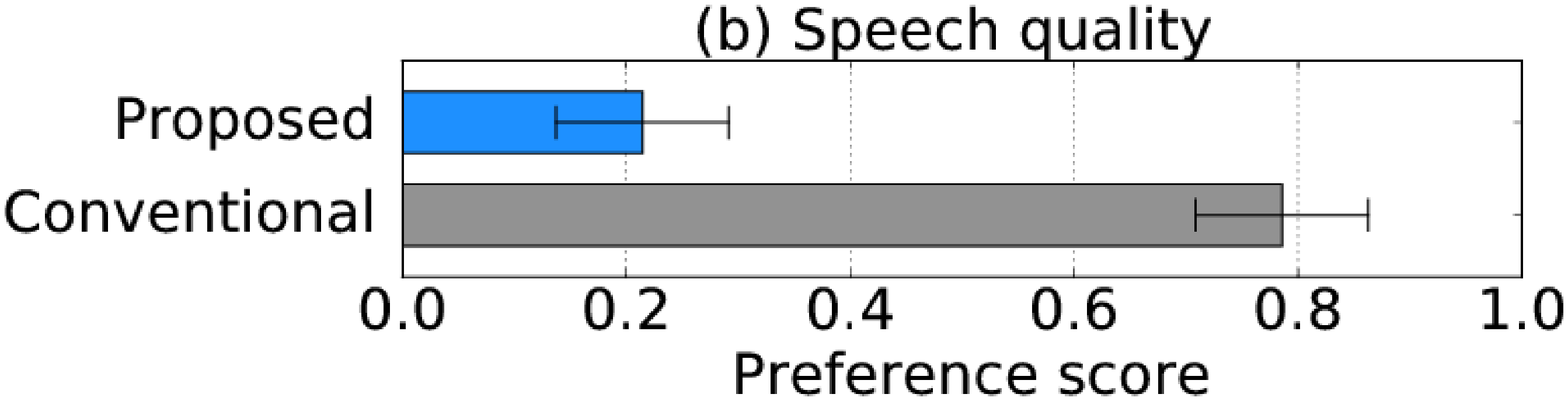}
  \caption{
    Results of subjective evaluation for comparing conventional VC
    and
    proposed VC integrating posterior probability conversion.
    Error bar indicates the 95~\% confidence intervals.
  }
  \label{syukan}
  \vspace{-20pt}
  \end{center}
\end{figure}

The results are shown in Fig.~\ref{syukan}.
Althogh the proposed VC performed better in speaker similarity
(Fig.~\ref{syukan}(a))
thanks to posterior probability conversion,
it degrades speech quality (Fig.~\ref{syukan}(b)).
It seems that the probability conversion caused conversion error
that missed the phonetic properties of the source speech parameters,
which is probably what resulted in the degraded quality.

\subsection{Evaluation of Joint Training}
\vspace{-2pt}
\subsubsection{Joint Training of Recognition and Synthesis}
\begin{figure}[t]
  \begin{center}
  \includegraphics[width=0.95\linewidth]{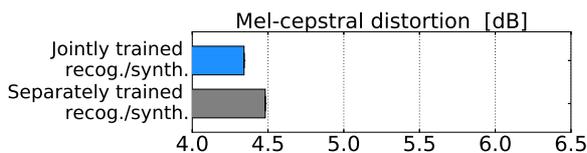}
  \caption{
    Mel-cepstral distortion in auto-encoding case.
    The conventional separately trained
    and
    proposed jointly trained
    recognition and synthesis modules are compared.
  }
  \label{comparison}
  \vspace{-15pt}
\end{center}
\end{figure}

We evaluated the effectiveness of joint training of recognition and synthesis modules,
in comparison to conventional separately trained modules~\cite{LSun16}.
We calculated mel-cepstral distortion in the auto-encoding case,
i.e.,
reconstruction error of source speech parameters
through recognition and synthesis.
As shown in Fig.~\ref{comparison},
the proposed joint training achieved better distortion than
conventional separated training.

\begin{figure}[b]
  \begin{center}
  \includegraphics[width=0.95\linewidth]{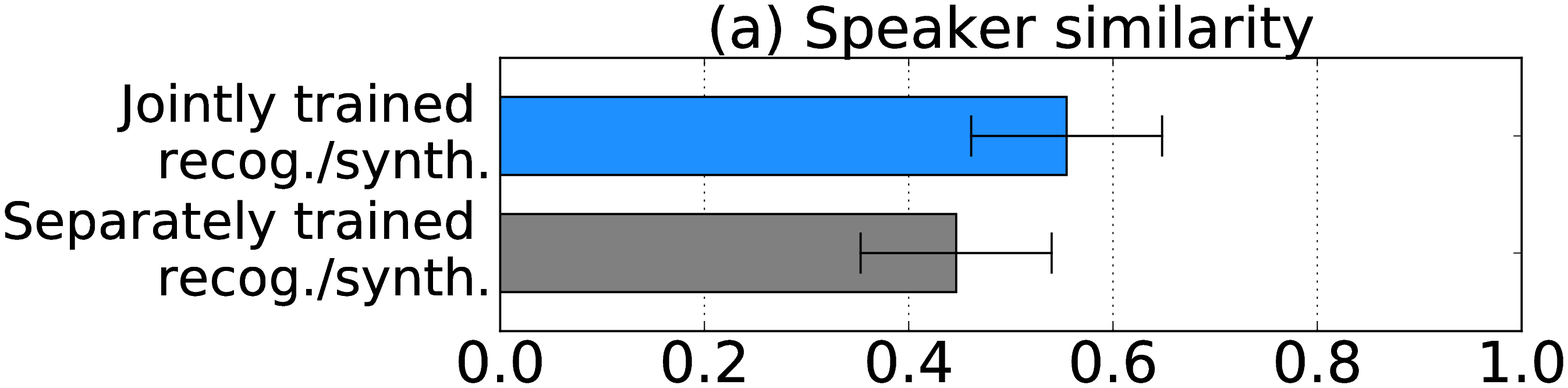}
  \vspace{5pt}

  \includegraphics[width=0.95\linewidth]{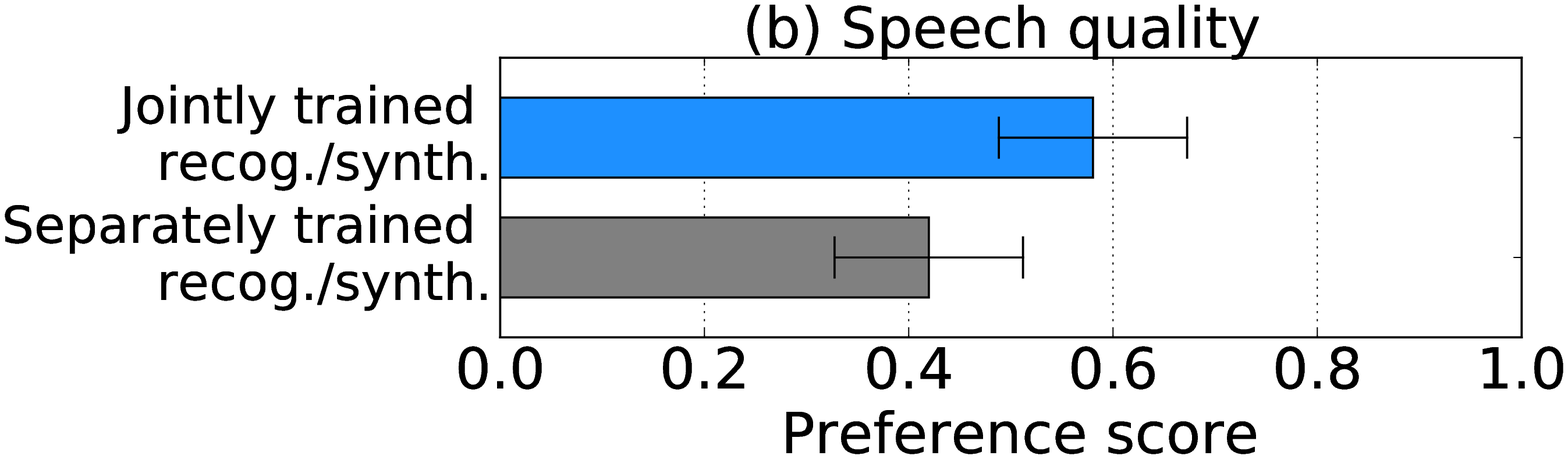}
  \caption{
    Results of objective evaluation for comparing conventional separately-trained
    and
    proposed jointly-trained recognition and synthesis modules.
    Error bars indicate the $95~\%$ confidence intervals.
  }
  \label{syukan2}
  \vspace{-20pt}
\end{center}
\end{figure}

We also performed an AB test on speech quality
and
XAB test on speaker similarity in the VC case,
as similar as in Section 4.2.2.
As shown in Fig.~\ref{syukan2},
the proposed joint-training overcomes the conventional separated
training in both speaker similarity and speech quality.

\subsubsection{Joint Training of Recognition, Conversion, and Synthesis}
To evaluate of the joint training of recognition,
conversion, and synthesis,
we calculated the mel-cepstral distortion of three systems:
(1) conventional VC~\cite{LSun16},
(2) separately trained modules (equal to "Proposed" in Section 4.2.1),
and
(3) jointly trained modules.
\begin{figure}[t]
\begin{center}
  \includegraphics[width=0.90\linewidth]{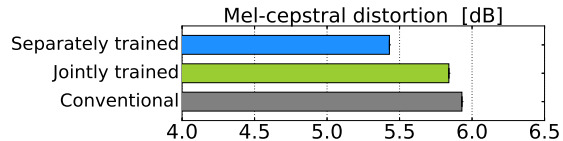}
  \caption{
    Mel-cepstral distortion of three methods:
    (1) conventional VC,
    (2) proposed VC using separately trained recognition/synthesis,
    and
    (3) proposed VC using jointly trained recognition/synthesis/conversion.
  }
  \label{joint_mcd}
  \vspace{-20pt}
\end{center}
\end{figure}

\begin{figure}[t]
  \begin{center}
  \includegraphics[width=0.99\linewidth]{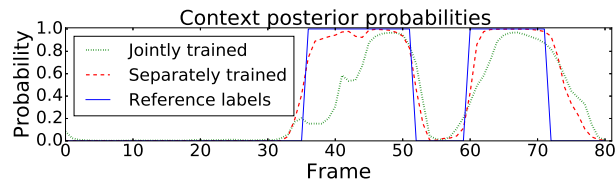}
  \caption{
    An example of posterior probability sequence.
  }
  \label{ppg_dif}
\end{center}
\end{figure}

The results are shown in Fig.~\ref{joint_mcd}.
The joint training scored better than conventional VC
but worse than separated training.
To clarify this,
we show in Fig.~\ref{ppg_dif} an example of
probability sequence estimated through speech recognition.
The separately trained recognition module outputs harder probabilities,
i.e.,
the values are close to 0 or 1 for all frames.
However,
we can see that the joint training of recognition and synthesis
tends to make the values soft.
This requires deeper investigation,
but we suspect this tendency is one of the reasons.

\vspace{-5pt}
\section{Conclusion}
In this paper,
we proposed voice conversion (VC) using sequence-to-sequence learning
of context posterior probabilities.
Since conventional VC directly uses the posterior probabilities of source speech
for predicting target speech,
it is difficult to convert the speaker individuality
included in the posterior probabilities.
To address this,
we built sequence-to-sequence conversion models
that convert the source context posterior probability sequence into a target one.
Further,
we proposed joint training algorithms for
speech recognition,
speech synthesis,
and
posterior probability conversion.
Experimental results demonstrated that
(1) the proposed algorithms outperformed
the conventional VC in speaker similarity,
and
(2) joint training of recognition and synthesis outperformed
the conventional VC in both speaker similarity and speech quality.
As future work,
we will investigate how to determine the frame length of the
converted posterior probability sequence.

{\bf Acknowledgements:}
Part of this work was supported by ImPACT Program of
Council for Science, Technology and Innovation
(Cabinet Ofﬁce, Government of Japan), SECOM Science and Technology
Foundation, and JSPS KAKENHI Grant Number 16H06681.

\bibliographystyle{IEEEtran}
\bibliography{bibtex_interspeech_rev}

\end{document}